Research Article for BMC Bioinformatics

# JAK/STAT signalling - an executable model assembled from molecule-centred modules demonstrating a module-oriented database concept for systems- and synthetic biology

Mary Ann Blätke[1], Anna Dittrich[2+], Christian Rohr[1,3+],

Monika Heiner[3§], Fred Schaper[2§], Wolfgang Marwan[1§*]

* Correspondence: wolfgang.marwan@ovgu.de

+ Contributed equally

§ Shared senior authorship

[1]Lehrstuhl für Regulationsbiologie and Magdeburg Centre for Systems Biology, Otto-von-Guericke Universität

## Abstract

We describe a molecule-oriented modelling approach based on a collection of Petri net models organized in the form of modules into a prototype database accessible through a web interface. The JAK/STAT signalling pathway with the extensive cross-talk of its components is selected as case study. Each Petri net module represents the reactions of an individual protein with its specific interaction partners. These Petri net modules are graphically displayed, can be executed individually, and allow the automatic composition into coherent models containing an arbitrary number of molecular species chosen *ad hoc* by the user. Each module contains metadata for documentation purposes and can be extended to a wiki-like minireview. The database can manage multiple versions of each module. It supports the curation, documentation, version control, and update of individual modules and the subsequent automatic composition of complex models, without requiring mathematical skills. Modules can be (semi-) automatically recombined according to user defined scenarios e.g. gene expression patterns in given cell types, under certain physiological conditions, or states of disease. Adding a localisation component to the module database would allow to simulate models with spatial resolution in the form of coloured Petri nets. As synthetic biology application we propose the fully automated generation of synthetic or synthetically rewired network models by composition of metadata-guided automatically modified modules representing altered protein binding sites. Petri nets composed from modules can be executed as ODE system, stochastic, hybrid, or merely qualitative models and exported in SMBL format.

## Background

Systems biology is a multidisciplinary approach to understand the complex molecular mechanisms of life by combining systems theory and experimental approaches. For experimentalists and theoreticians it is not easy to find a common language and this is still a bottleneck for successful project development and cooperation. Although systems biology has developed into an established discipline, the majority of life scientists is still not aware of the benefits of kinetic models. Some scientists are even skeptical about the usefulness of a model based systems view on biological mechanisms. One obvious reason for this situation might be that life scientists working as experimentalists are traditionally not (well) trained in mathematics, modelling or even systems theory. This bars them from being able to access, use, and judge kinetic models of molecular networks. Furthermore, suitable platforms designed to facilitate productive interactions between experimentalists and theoreticians are widely missing but would allow more experimentalists to contribute to gain a systems view on biology.

Petri nets are a formal modelling language which, according to our experience, is quickly learned and intuitively understood by pure experimentalists as the graphical representation and execution of a Petri net is similar to the way of how molecular reaction schemes are usually graphically displayed in the field of biochemistry and molecular biology (Figure 1). In Petri nets, molecular components are represented as places and biochemical reactions as transitions [1]. Accordingly, the translation of a biochemical reaction scheme into a Petri net model is straightforward (Figure 1). A Petri net can be interpreted as a qualitative or a quantitative (ODE, stochastic, or hybrid) model while the graphical representation of its structure looks identical.

Appropriate tools make Petri nets powerful executable models [2] where the equations required for running simulations are automatically generated in the background while the model is drawn in a graphical user surface (Figure 1 E,F). Petri net tools such as Snoopy provide a unifying framework for the graphical display, computational modelling, and simulation of molecular networks [3, 4]. In Snoopy, a Petri net graph can be interpreted as a qualitative or a quantitative model that allows to perform continuous (ODE-based) or stochastic simulations and even hybrid approaches [5]. In addition, colored Petri nets permit to simulate heterogeneous populations of interacting molecules or populations of cells by automatically creating multiple, eventually interconnected copies of a Petri net and by running them in parallel [6]. Snoopy supports the export of models in the form of ODE's or as SBML code in order to support the exchange with existing modelling and simulation platforms. Together with the analysis tool Charlie and the model checking tool Marcie [7], Snoopy contributes to a computational platform supporting advanced Petri nets for systems biology and provides the basis for the work described here.

Kinetic models of molecular networks common to systems biology are usually monolithic entities describing the functional interactions of a certain number of molecular components. Up to now many models have been built based on both literature data and experimental results. For various reasons, different models often cannot be easily combined to more comprehensive models that still perform correctly as the combined parts do on their own (see Discussion). Models describing the same biological process may, for example, differ considerably due to a distinct degree of details. Therefore enlarging as well as updating of existing models is a difficult, time-consuming, and an error-prone task.

We propose that computational models of molecular networks should be semi-automatically generated or updated from a central collection of curated modules, each reflecting the functional interactions of an individual molecule such as a specific protein or RNA. We will discuss functional features to make such collections useful for the expert user and the non-expert user as well.

Taking the JAK/STAT signalling pathway as example, we assemble an executable model from a collection of Petri net encoded modules organized in database prototype.

The JAK/STAT pathway is a prevalent intracellular signalling pathway activated by a multitude of cytokines such as interleukins, interferons, and growth factors [8, 9]. The canonical JAK/STAT pathway is activated by binding of a ligand to membrane bound receptors and subsequent phosphorylation and activation of receptor-associated JAKs. Activated JAKs phosphorylate tyrosine residues in the cytoplasmic part of the receptors, which serve as binding sites for STATs. After binding, STATs are phosphorylated by JAKs, dimerize, and translocate to the nucleus where STATs act as transcription factors (Figure 2). The activation of the JAK/STAT pathway is restricted by the transcriptional induction of feedback inhibitors e.g. SOCS3 and constitutively expressed phosphatases e.g. SHP2. The JAK/STAT signalling pathway can be activated in a variety of different cell types and through a plethora of receptors resulting in the regulation of different sets of genes. Thus, JAK/STAT signalling has a broad range of effects. This results in a complex network of signalling events that are not yet fully understood.

JAK/STAT signalling plays a major role in development, cell proliferation, cell migration, and inflammatory processes [10]. Dysregulation, especially constitutive activation, of JAK/STAT signalling was found in cancerous, autoimmune, and inflammatory diseases. Here we refer to the IL-6 type cytokine-induced JAK/STAT pathway activated e.g. by IL-11, LIF (leukemia inhibitory factor) and IL-6 itself. IL-6 is a crucial regulator of both pro- and anti-inflammatory processes [11] and dysregulated IL-6 signalling is associated with e.g. rheumatoid arthritis, multiple sclerosis and plasmacytoma [9].

The combinatorial effects of different IL-6-type cytokines and their receptors and the involved isoforms of signalling molecules of the JAK/STAT pathway exemplify the usefulness of the modular modelling concept described here. All signalling molecules are implemented individually as independent model modules in the form of a Petri net equipped with connection interfaces representing the kinetic mechanisms of interaction with other components. These modules can be recombined automatically in a combinatorial manner to generate executable kinetic models for the different scenarios of JAK/STAT signalling.

## Results and Discussion

### General description of the modular modelling concept

In its basic form, the modular modelling concept considers every protein as functionally independent model, here called module. A module considers the (1) reaction mechanisms and (2) molecular interactions of a given protein with other proteins or low molecular weight molecules (e.g. 2nd messengers). Furthermore, a module includes (3) conformational changes affecting functionally relevant states, and (4) the binding of and dissociation from other proteins or other molecules (e.g. lipids). A module also comprises (5) all known biochemical modifications that might influence the activity of the protein or the probability of its interaction with proteins or other molecules, e.g. posttranslational modifications. Technically, each module is organized in the form of hierarchically structured submodels in Snoopy (see Figure 1 C,D for the explanation of submodels and for further details). Where applicable, small molecules like ions, second messengers, energy carriers, metabolites etc. are included as intrinsic components of a module such that two or more proteins may functionally couple with each other through their interactions with the same (small) molecule.

We found that each module exhibits certain structural properties that influence the dynamic behaviour of the entire module as listed in Table 1 and described in [12]. These properties can be used as criteria to validate the functional integrity of each module. Modules validated accordingly can be automatically linked through common parts, called common subnetworks to gain a modular network (Figures 3,5,6; for details see below). From the technical point of view it is required that the places and transitions of the subnetworks of a module are declared as logical nodes (Figures 3,5) which ensure that the marking of the places shared between modules is the same in the connected modules. Interfaces between given modules representing interacting molecules and their interaction mechanism are defined by interface subnetworks. We use the terms interface subnetwork and submodel differently. According to our definition, an interface subnetwork connects modules while a submodel refers to the hierarchical representation of parts of a network within Snoopy. Interface subnetworks may be represented as submodels in Snoopy but by far not all submodels are interface subnetworks.

We also found that one can predict the properties of a automatically composed model based on the shared topological properties of the individual modules (Table 1). These topological properties as revealed by the analysis tool Charlie [12] can be understood in terms of biologically relevant molecular mechanisms [13]. Composition by automatic coupling does neither alter the structure nor the properties of each module, but defines the structural properties of the composed model. Any automatically composed model is fully executable and may be interpreted as a qualitative, stochastic, continuous, or hybrid model for running simulations (see Figure 1 for details). In the following, we will detail the procedure of constructing a protein module.

**Constructing a module**

Modules are constructed based on the domain structure and molecular reaction mechanisms of a protein as well as its interactions with other molecules. This includes also conformational states that are functionally relevant. For each module, literature references are annotated as part of the module's database entry. Each place of a module corresponds to a specific functional state of a specific protein domain (e.g. a phosphorylated or unphosphorylated side chain, a catalytically active or inactive domain etc.). A place may also represent a specific status of a protein with respect to the interaction or reaction with a small molecule (e.g. a binding site free or occupied by a small molecule, substrate, or product etc.). In this context, a transition as node of the Petri net represents a molecular transition between two different states of a protein domain, any biochemical reaction of the protein, or the physical interaction of the protein (domain) with other molecules or domains (Figure 1). Each module fulfills the structural properties given in Table 1.

As protein modules in general describe the detailed reaction mechanism of a protein in terms of its catalytic and regulatory reaction cycles, the underlying Petri net reflects mass conservation with respect to the modeled protein. Mass conservation results in the coverage of any protein module by one or several P-invariants. The minimal P-invariants describe the set of all possible states of a protein domain. These states may be involved in the interaction with small molecules or with a second site of the same or of an interacting protein (for a brief introduction to P- and T-invariants see the extended version of legend to Figure 1 provided as supplemental file 1). Protein modules do always contain T-invariants. A minimal T-invariant consists of a set of transitions that restores the initial marking of a subnet. The minimal P-invariants of a module, on the other hand, can be classified as strongly connected (closed) or not strongly connected (open) P-invariants. Both types constitute a state machine (a net without forward and backward branching transitions). A strongly connected state machine (a cycle) corresponds always to a T-invariant, which may or may not be minimal. Thus, closed P-invariants are covered by T-invariants and *vice versa*. Accordingly, the initial marking of a closed P-invariant representing the reaction cycle of a protein domain can be restored and the corresponding part of the Petri net hence is live. This is not true for a domain that is represented by an open P-invariant. Those parts of a protein module that represent the interaction with small molecules form an open P-invariant, and the corresponding net may run into a dead state if the supply with a small molecule is exhausted.

Places representing molecular states that catalyse a reaction are connected to the respective transition by a double arc meaning that a token respresenting the catalyst is consumed and restored when the catalytic reaction occurs. To avoid external sinks and sources for protein domains, a protein module is never bordered by transitions. Instead, a protein module is always bordered by places. However, boundary places may correspond to places in modules of other proteins or may represent small molecules or other chemical components either consumed or produced.

The principle of double entry bookkeeping is a hallmark and a necessary consequence of the modular modelling concept, since every module must represent all of its interactions with other molecules. Thus, modules of two interacting proteins share exactly matching interface subnetworks describing the mechanism of interaction (Figure 3). In Snoopy, the interface subnetworks may be implemented in the form of a submodel (Figure 1D; Figure 3).

The described approach to generate modular, compositional models of proteins can be extended by designing modules that model the synthesis and degradation of RNAs and proteins. This leads to two distinct classes of modules with distinct structural properties: (1) protein modules describing functional interactions of proteins and (2) biosynthesis/degradation modules describing the synthesis, degradation, and hence stability of proteins or RNAs. By including biosynthesis/degradation modules one can model the regulation of gene expression and translation, but also e.g. the formation of non-coding RNAs and their regulatory influence on protein biosynthesis. Protein modules are covered with P-invariants. In contrast, biosynthesis/degradation modules are covered by T-invariants.

**The JAK/STAT signalling pathway assembled from Petri net encoded modules**

In the following, we will first describe why and how the JAK/STAT signalling pathway was organized into modules. We will explain the principles underlying the structural organization and validation by taking the gp130 module as an example. We then demonstrate how the modules were assembled to generate coherent models of the JAK/STAT pathway. We will perform structural analyses, and run simulations to reveal model discrimination for alternative models of JAK activation. Finally, we will describe the design of the database for organizing modules and the composition of the modules to executable models.

Signal transduction through the JAK/STAT pathway is induced by many interleukins and all interferons. A limited number of specific JAK kinases, STAT transcription factors and receptor subunits leads the redundant usage of specific signalling components in response to specific stimuli, e.g. the different IL-6-type cytokines signal all through receptor complexes containing at least one gp130 receptor chain. Nevertheless, the combination of individual signalling components activated in a certain cell type by a well-defined stimulus leads to a specific biological response. However, using identical or similar signalling molecules allows complex cross talk mechanisms between individual signalling cascades. Thus, the information flux through the JAK/STAT network and hence the resulting dynamical behavior of the system depends on the stimulus, the cell type-specific expression profiles of the individual signalling components and may also change depending on the specific physiological conditions. Thus, models may have to be adjusted according to the components that make up the network structure in order to obtain realistic simulations.

To compose executable models of the JAK/STAT signalling pathway, we designed a separate Petri net module for each individual protein including the specific molecular interactions with other components of the pathway. All modules were organized into a database from which they can be recombined. The modules are automatically connected through logical places which represent shared molecular components or states resulting in a single, coherent Petri net model of the entire pathway.

In the example shown in Figure 4, we consider the combinatorial variety, which is given by the interactions of three cytokines (IL-6, IL-11, and LIF), three cytokine receptors (IL-6R$\alpha$, IL-11R$\alpha$ and LIF-R), the common receptor chain gp130 and the two signalling proteins JAK and SHP2. The involved components are represented by specific modules. One of these modules (gp130) will be explained in more detail below. To allow for the discrimination of competing mechanisms, two concurrent modules accounting for the alternative molecular reaction mechanisms have been derived for both JAK and SHP2. To couple the modules in all possible combinations, all potential interactions represented by coarse transitions had to be declared as common network motifs, as illustrated in Figure 3. In case of the gp130 module, five coarse transitions represent possible (i.e. experimentally confirmed) interactions with other signalling components (Figure 4). Competing functionally correct and executable models can be generated by coupling appropriate modules in arbitrary combinations. Unused coarse transitions are inactive, because the corresponding subnet by default is not sufficiently marked (Figure 5) while it automatically becomes sufficiently marked when the two interacting modules are composed. The coarse transitions in a module constitute subnets for all possible interactions with different proteins and other molecules. This ensures compositionality of all modules without the need to adjust their structure for assembling a composed model.

## The gp130 module as a representative example for a modular Petri net: design principles, structural features and validation

To explain the composition of an exemplary single module, we describe the structure of gp130. All other modules listed in Table 2 are given as graphical representation and in the form of an executable Snoopy file as supplementary material. The Petri net module of gp130 has been constructed according to the functional and structural properties of the protein (ligand binding site, binding sites for JAK1 and phosphotyrosine motifs representing binding sites for SHP2, STAT3 and SOCS3). For the sake of a clearly readable model structure, we represent the crucial functional entities of the gp130 by coarse transitions (Figure 6). Each phosphotyrosine (see also Figure 2) is connected to at least three coarse transitions, one describing the phosphorylation, another describing the dephosphorylation of the receptor motif and the remaining coarse transitions describe the binding of an interaction partner to the respective phosphotyrosine motif. Phosphorylation and dephosphorylation of the receptor motif, as well as binding of a specific signalling component occur in principle through the same mechanism for each phosphotyrosine motif within gp130. Because of these recurring motifs, the network structures describing the detailed kinetic mechanisms are represented as submodels of the respective coarse transitions (Figure 6, red boxes). In general, hiding recurring mechanisms and kinetic details by coarse elements provides the following advantages: (1) Clearly defined submodels can be easily cloned via copy-paste. (2) Hiding recurring mechanisms and kinetic details stress the focus on functionally important elements of the network and on their functional interactions in terms of regulatory mechanisms. The interactions between functionally coupled sites of a protein (resulting in e.g., allosteric effects; binding sites for an enzyme and its substrates) become clearly visible in the form of arcs connecting the different functional parts within the Petri net representation. The degree of interaction given by this representation gives a visual impression of the functional sites where conformational changes are triggered and the activity and functionality of the respective protein is regulated (Figure 6).

After constructing the gp130 module, it was tested for functional integrity by analyzing the corresponding Petri net in Charlie [12]. The analysis of the module confirmed the expected structural properties as given in Table 1. Finally, the predicted dynamic behavior of the module was analysed by running simulations (not shown).

**Composing and testing the modular network**

To assemble the modular networksignalling, the following steps were taken: (1) subnets of coarse transitions describing interactions with other components are declared as shared subnets via logic places and transitions, (2) each module is wrapped into a coarse place and (3) copied into one Snoopy file holding the modular network (in a future version of the database the composition will be performed fully automatically). After generating the entire modular network, the Petri net was tested for structural integrity using Charlie. Again, the expected structural properties for the composite network were confirmed.

The model of IL-6-induced Jak/STAT signalling was assembled accordingly from the set of modules listed in Table 2 and tested against experimental data (Figure 7). The model comprises the stepwise binding of IL-6 to its two receptors subunits IL-6R and gp130 and the subsequent activation of JAKs and its feed-back control. We analyzed if the model outputs are able to reflect experimental data.

Therefore, HEK293 cells stably expressing gp80 were stimulated with a 6 minutes pulse of IL-6. To analyze the initial activation of the JAK/STAT pathway in detail the cells were harvested in one minute time intervals. The dynamics of JAK1, STAT3 and SHP2 activation were measured by quantitative immunoblotting as described elsewhere (Dittrich et al., 2012, submitted). All three species are not phosphorylated in the absence of IL-6. Stimulation with IL-6 induces a fast increase of phosphorylation of JAK1, STAT3 and SHP2 within minutes. About one minute after the initial stimulus has been removed by washing the cells, the phosphorylation of the three proteins reaches a plateau and remains constant within the experimentally analyzed time window. The model output qualitatively matches the experimental data with respect to both, the stimulus-induced phosphorylation of JAK1, STAT3 and SHP2 and the plateaus reached after removal of the IL-6 stimulus (Figure 7). The concept of modular organization of networks offers a tool to expert modelers as well as to non-experts to easily test the influence of locally alternative molecular mechanisms on the behavior of the entire system.

**Structure of the module database**

We implemented a first version of a database for the organisation, handling, and composition of Petri net modules and a web-interface to provide public access. In its current version, the database allows to browse through the collection of modules while the automatic composition of networks from modules is not implemented yet. The relational database management system used is MySQL. We organized the database in four parts holding information about (1) modules, (2) proteins, (3) used literature and (4) users (Figure 8). The core part of the database is the module scheme. This scheme holds tables describing all relevant information about each module and tables connecting the other components of the database. Information about the proteins represented, as modules are stored in the protein scheme, including e.g. protein names, gene symbols, accession numbers, short names, alternative names and relevant pathways. This information has been extracted from the UniProt database via the assignment of the accession number of the proteins modeled. The literature scheme contains bibliographic details of publications that have been used for the design of the respective module and links via PMID to PubMed [14]. The user scheme holds contact and login information of the user. The structure of the database is easily extensible to adopt features to be implemented in the near future and also to adapt to further user needs. The system is able to query for different entities (modules, places, transitions, etc.) according to ambiguous search terms like name or class of a protein, a gene symbol, an accession number, or an author, etc.

**Composing models by browsing through the database**

While browsing through the database, the user is systematically guided through the collection of available modules (Figure 9). Users can compose coherent models from modules of interest interactively. Composition of an entire model may be performed without directly touching any of the modules selected. Instead, composition of models is solely based on the information provided in the table view of the database (Figure 9). Viewing the graphical representation of the individual Petri net modules is possible but not even necessary. Upon command, the executable coherent model is composed automatically for structural analysis or simulation (this feature has not been implemented in the prototype). The graphical representation of the coherent Petri can be viewed along with the documentation of its modular components.

By entering a search term like „STAT“ into the browser window of the web interface of the database (Figure 9), one receives a list of all modules that contain places representing any kind of the STAT transcription factor as component. The list includes all modules for molecules that interact with any type/state of the JAK1 protein. The user may select a specific module in the displayed list, e.g. SOCS3 to obtain a new list of all modules that are functionally coupled to SOCS3. Subsequently, a selection of modules (one of which represents gp130) can be chosen for building up a coherent model by marking the appropriate check boxes. Clicking the names of a checked module leads to a corresponding list of interconnected modules that may be selected for being included in a composed model, too.

Alternatively, one may enter the search term for a specific organism into the browser window in the Advanced Search view to receive a list of all modules that model proteins of e.g. human cells and create composed models as described. A composition of modules across species is not supported according to the default settings of the modules because the specific interaction sites between molecules may differ between organisms (e.g. different serine residues may be phosphorylated in the mouse as compared to the human protein) and thus will require specific Petri net submodels for coupling of the modules. Nevertheless, cross species connections for synthetic biology purposes can be introduced by defining and editing a new version of a module which specifies the respective interaction, e.g. the phosphorylation of a specific tyrosine side chain in a human receptor by a murine kinase.

Each module contains its specific documentation in the form of name, version number, link to the Uniprot accession number of the corresponding protein, organism and cell type the module refers to, relevant Literature, supplementary information, names of authors and curators, link to previous and alternative versions of the module, and date of the last update. It also contains links to versions of the module that might have been obtained by model reduction and *vice versa*. If authors from different laboratories disagree on mechanistic details, they might be willing to supply alternative versions of a module for the database along with information supporting the proposed alternative mechanisms. This might be particularly useful for proteins analyzed in different cell types or for proteins obtained by alternative splicing.

**General benefits of the modular modelling approach for biomodel engineering**

In contrast to a collection of monolithic models, the database established here contains exclusively Petri nets in the form of modules that can be assembled to coherent models. Working with modules seems unnecessarily complicated at first. However in contrast to conventional monolithic models, the strict modularity together with the organization of the components of all modules within one searchable, relational database provides several important advantages and options.

In addition to convenient management and easy recombination of modules within the database, the modular modelling approach provides a variety of benefits for expert and non-expert users and encourages the development of automated approaches to biomodel engineering:

**I. Modules are reusable and exchangeable and contain metadata**. Besides its functionality as an exchangable component of a coherent Petri net model, a module serves as a comprehensive graphical, wiki-like review on the functional interactions and molecular properties of the protein it represents. The metadata are meant to document any assumptions, simplifications, and controversies regarding the represented molecular mechanisms and automatically become part of any composed model. For example, composed models can be easily queried for the latest update of each of its components.

Modules can be incorporated into different models without laborious and error-prone reorganisation and adaptation (Figure 10). To analyse the functional consequences of alternative mechanisms for example, one can analyze how the performance of the model changes upon exchanging certain modules by versions considering alternative mechanisms or by versions of a module obtained through methods of model reduction.

Creating different versions of a module (according to alternative molecular mechanisms) from the database should be made possible for all users. However, the permission to edit publically available individual modules should be restricted to authorized users to prevent uncontrolled editing.

**II. Modules facilitate expert curation and updating by experimentalists from life sciences***.* Many scientists in life sciences have deep knowledge on a specific protein (or family of proteins) and its interaction partners, biochemical properties and molecular functions. For these experts it is presumably easier and more straightforward to survey and update an isolated module rather than to identify and to work on the corresponding relevant parts in the context of a complex monolithic model. Furthermore, the modular concept prevents divergence into a variety of monolithic model versions that *post synthesis* are difficult to be judged with respect to the molecular mechanisms they implement and the assumptions and simplifications they make. Modules in a database encourage researches to keep an eye on their favorite molecules.

**III. Models can be composed according to gene expression patterns of certain cell types or under certain physiological (or pathological) conditions***.* The modular approach facilitates the incorporation of specifically those modules that correspond to these genes expressed in the specific cell of interest, e.g. the expression of the receptor subunit gp80 of the IL-6-receptor complex is restricted to a low number of cell types (e.g. hepatozytes or leukocytes). All possible connections of a specific module are activated during the automatic composition while all unused connections remain inactive because the respective places do not contain any tokens (Figure 5).

**Simulation of biochemical reactions in a reaction-diffusion system with the help of a localisation component.**

Adding colour to tokens enables us to consider the spatial distribution of events where same colours indicate identical localisation. Only components residing in the same compartment are able to interact. In coloured Petri nets, the firing of transitions may depend on the colour of tokens and tokens may change their colour through the firing of transitions in a freely defined way. Coloured Petri nets as implemented in Snoopy therefore combine the formalism of Petri nets with the expressive power of a computer language [6]. Entering the world of coloured Petri nets, the consistently modular design of our approach can be extended to allow the simulation of biochemical reactions in a reaction-diffusion system with the help of a localisation component. The localisation component, together with the other modules, may be automatically assembled to generate any composed model with spatial resolution. As all other modules, the localisation component may be provided in different versions. In its most simple form, the location module may just indicate the cellular compartment in which a biomolecule resides (Figure 11A). The STAT dimer, for example, can only bind to DNA if the STAT has translocated from the cytoplasm into the nucleus. The actual localisation of a molecule in any compartment of the cell can be encoded by the color of the token that marks a place representing the given molecule. The logical transitions of a submodel defining the interface between two modules are set to fire only if the tokens in their pre-places are of the same color (Figure 11D). Different colors indicate that two molecules are in different compartments and accordingly cannot react with each other.

Reaction-diffusion systems with arbitrary spatial resolution may be modeled with a version of the localisation component consisting of a three-dimensional lattice of places that represent corresponding elements of the reaction volume (Figure 11B). While moving through the lattice by firing of randomly chosen transitions which connect adjacent places (Figure 11C), the tokens change their colors according to their current position. The transitions must be programmed differently in the localisation component as compared to the protein modules and the biosynthesis/degradation modules. In the localisation component, the transitions change the colour of the tokens which they switch according to a pre-defined neighbouring rule. In the protein modules and in the biosynthesis/degradation modules, the transitions of the interfaces are selective with respect to the colours of the tokens they switch. However, firing of transitions does not change the color of the switched tokens.

By defining the token colour as a tuple with four entries (Figure 10D), one may assign the localisation in terms of the cellular compartment to each reaction volume. In other words, the three-dimensional lattice (Figure 11B) may be fitted with a three-dimensional topological model of the entire cell (Figure 11A). Faithful 3-D models can be obtained by confocal microscopy or electron tomography [15] for example. Like all modules in the database, the localisation modules can be reused in conjunction with the different biochemical models which keeps the programming effort to a minimum.

Following the described concept, the user will not get in direct touch with the unfolded Petri nets or with the programming of colored Petri nets by employing the localisation component. These nets might be generated automatically through algorithms during the assembly of the composed models while the user only needs to interact at the level of the Tables View of the database. The algorithmic realisation may rely on the Petri net in the Petri net concept where each token in the Petri net representing the localisation component is a Petri net composed of modules which represents the dynamically interacting biomolecules. Regarding the module database this means that the modules remain the same no matter whether they will be used for composing a non-spatial model or for a model which considers cellular compartments and spatial resolution. The modules are only interpreted as coloured Petri nets if a space module is used.

**Design of synthetic or synthetically rewired networks by combination of modules representing altered binding sites**

Composing computational models from modules with altered binding sites allows to systematically change the structure of a network and to search for networks with desired behaviour. One visionary scenario of applying the module database to synthetic biology concerns the rational synthetic rewiring of preexisting networks through selectively altered binding sites of proteins. The idea is illustrated in Figure 12. Let us assume that two naturally occurring networks use two different kinases, Kinase 1 and Kinase 2, each phosphorylating three different substrates and that there is hardly any cross talk between the two kinases. By engineering Kinase 2, for example, its substrate specificity might be changed to functionally couple the two networks resulting in an altered system behaviour. Designing the new network and exploring its predicted function *in silico* may be performed interactively through structural alteration of e.g. the Kinase 2 module and reassembling the model accordingly.

In principle, reengineering modules by deleting existing and introducing new binding sites specific to other modules could be performed fully automatically and systematically while making essential use of the metadata assigned to each module in the database in order to adhere to biochemically realistic scenarios. *In silico* networks with alternative wiring could be assembled, again automatically, from re-engineered (mutated) modules and their performance queried for pre-defined properties. This systematic exploration of the functional potential of reengineered networks could make use of the tremendous computer power of the future. Presumably, correspondingly systematic but merely experimental approaches will remain out of reach simply because of the tremendous amount of work they would take.

The proposed *in silico* approach in fact would complement experimental random mutagenesis screens which often reveal phenotypes that are hard to fully understand. Moreover, for ethical reasons experimental mutagenesis will remain restricted to model organisms and is - in the case of multicellular organisms - further narrowed by the phenomenon of embryonic lethality. Hence it seems that exhaustive rewiring of networks *in silico* would definitely make sense.

We conclude that the described modular modelling approach represents a convenient and straightforward tool to generate dynamic and spatial models of biological processes for systems and synthetic biology.

## Competing interests

The authors declare that they do not have any competing interestes.

## Author's contributions

MAB developed and implemented the modular modelling approach, created the Petri net model of the JAK/STAT pathway, and implemented the web interface of the Petri net module database, with supervision by WM. AD and FS provided essential input and expertise for establishing the model of the JAK/STAT pathway. AD performed experiments to validate the model outputs. CR and MH made essential contributions in continuously developing Snoopy for biomodel engineering and in developing the web interface for the graphical display and animation of Petri nets. CR and MH also contributed ideas and technical advice for the organization and operation of a robust publically accessible database and for the remote automatic composition of executable models from modules. WM and MAB wrote the paper and MAB designed the figures. MAB and WM jointly conceived the study and worked together in its design and coordination. All authors read and approved the final manuscript.

## Author details

[1]Lehrstuhl für Regulationsbiologie and Magdeburg Centre for Systems Biology, Otto-von-Guericke Universität

[2]Lehrstuhl für Systembiologie and Magdeburg Centre for Systems Biology, Otto-von-Guericke Universität

[3]Chair of Data Structures and Software Dependability, Brandenburg Technical University Cottbus

## Additional Material

**Additional file 1: Extended Legend to Figure 1.**

**Additional file 2: Petri net models described in this work.** Unpacking of the ZIP file reveals XML files for the modules and the composed models listed in Table 2. The files can be opened and executed in Snoopy.

**Additional file 3: Chart of the JAK/STAT pathway in the form of Petri net modules.** The file provides the graphically displayed modules arranged according to the scheme of Figure 2.

## Acknowledgemnts

We thank Professor David Gilbert for critical reading of the manuscript and for suggesting module based automatic biomodel engineering and its application to synthetic biology. MAB and CR were financially supported by the International Max Planck Research School, Magdeburg.

## Websites

Snoopy:
http://www-dssz.informatik.tu-cottbus.de/DSSZ/Software/Snoopy

Charlie:
http://www-dssz.informatik.tu-cottbus.de/DSSZ/Software/Charlie

Marcie:
http://www-dssz.informatik.tu-cottbus.de/DSSZ/Software/Marcie

MySQL - The world's most popular open source database
http://www.mysql.com/

Protbricks:
http://www.protbricks.de

**Table 1 Topological properties of the modules and their biological interpretation.** The properties that must be fulfilled and those that must not be fulfilled equally apply to individual modules as well as to models composed from modules. For details see [12].

| Structural properties of modules that | | |
|---|---|---|
| **must be fulfilled** | **must not be fulfilled** | **may vary between modules** |
| Ordinary | Pure | Dead Transition |
| Homogeneous | Boundary Transitions | Dead states |
| Connected | Conservative | Dynamic conflict free |
| Covered with P-invariants | Static conflict free | Boundary places |
| Boundedness | Strongly covered with T-Invariants | Strongly connected |
| | | Non-blocking multiplicity |
| | | Covered with T-Invariants |
| | | Siphon-Trap Property |
| | | Liveness |

### Table 2 Overview modules and composed modular networks in numbers

| Module | Number of Places | Number of Transitions |
|---|---|---|
| IL-6 | 12 | 6 |
| IL-6R | 12 | 6 |
| gp130 | 53 | 48 |
| JAK1_V1 | 61 | 46 |
| JAK1_V2(1) | 61 | 47 |
| STAT3 | 24 | 21 |
| SHP2_V1 | 45 | 33 |
| SHP2_V2(2) | 49 | 41 |
| SOCS3 | 6 | 4 |
| SOCS3_Biosynthesis | 11 | 6 |
| Model 1 (with JAK1_V1, SHP2_V1) | 88 | 95 |
| Model 2 (with JAK1_V2, SHP2_V2) | 92 | 102 |

(1) Basal activity of JAK1
(2) Basal repressor

## Figures and Figure Legends

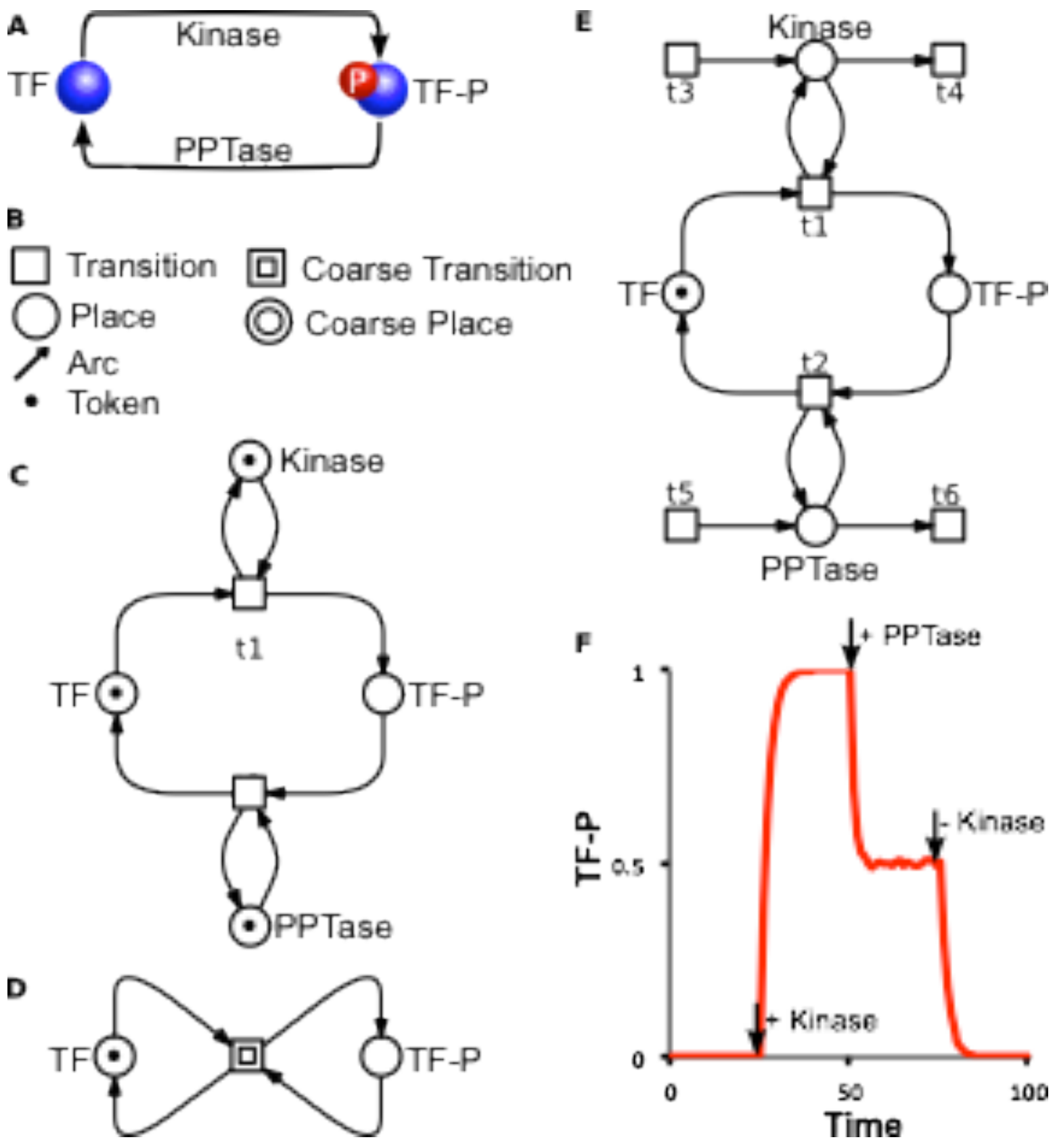

**Figure 1 Petri nets as executable models of biochemical reactions.** Petri nets are an intuitively understandable formal language to accurately describe biochemical reactions according to their molecular mechanism. Using the reversible phosphorylation of a protein as example, this figure briefly explains the Petri net formalism and gives an example of an executable model. (A) Biochemical reaction scheme of the phosphorylation of a transcription factor (TF) by a protein kinase and dephosphorylation of its phosphorylated form (TF-P) by a protein phosphatase. (B) Graphical elements of a Petri net and used to model biochemical reactions. (C) Petri net model of the reversible phosphorylation reaction shown in (A). Note that the model neglects formation and decay of enzyme-substrate complexes and hence will not show the effect of substrate saturation. (D) Coarse places and coarse transitions may be used to organize a Petri net into submodels. In the Petri net of panel (D), the kinase and the phosphatase reactions of (B) are lumped into a coarse transition. In Snoopy, coarse transitions represent submodels that are bordered by transitions whereas submodels represented by a coarse place are bordered by places. (E) The formation and decay and also the experimental addition or removal of biochemical components can be modeled by transitions [3, 16]. The kinetics in (F) was obtained by running (executing) the Petri net of (E) directily in Snoopy as a stochastic Petri net. A full-length version of this legend providing sufficient information for redears not familiar with Petri nets to essentially follow the present work is given as Supplemental File 1.

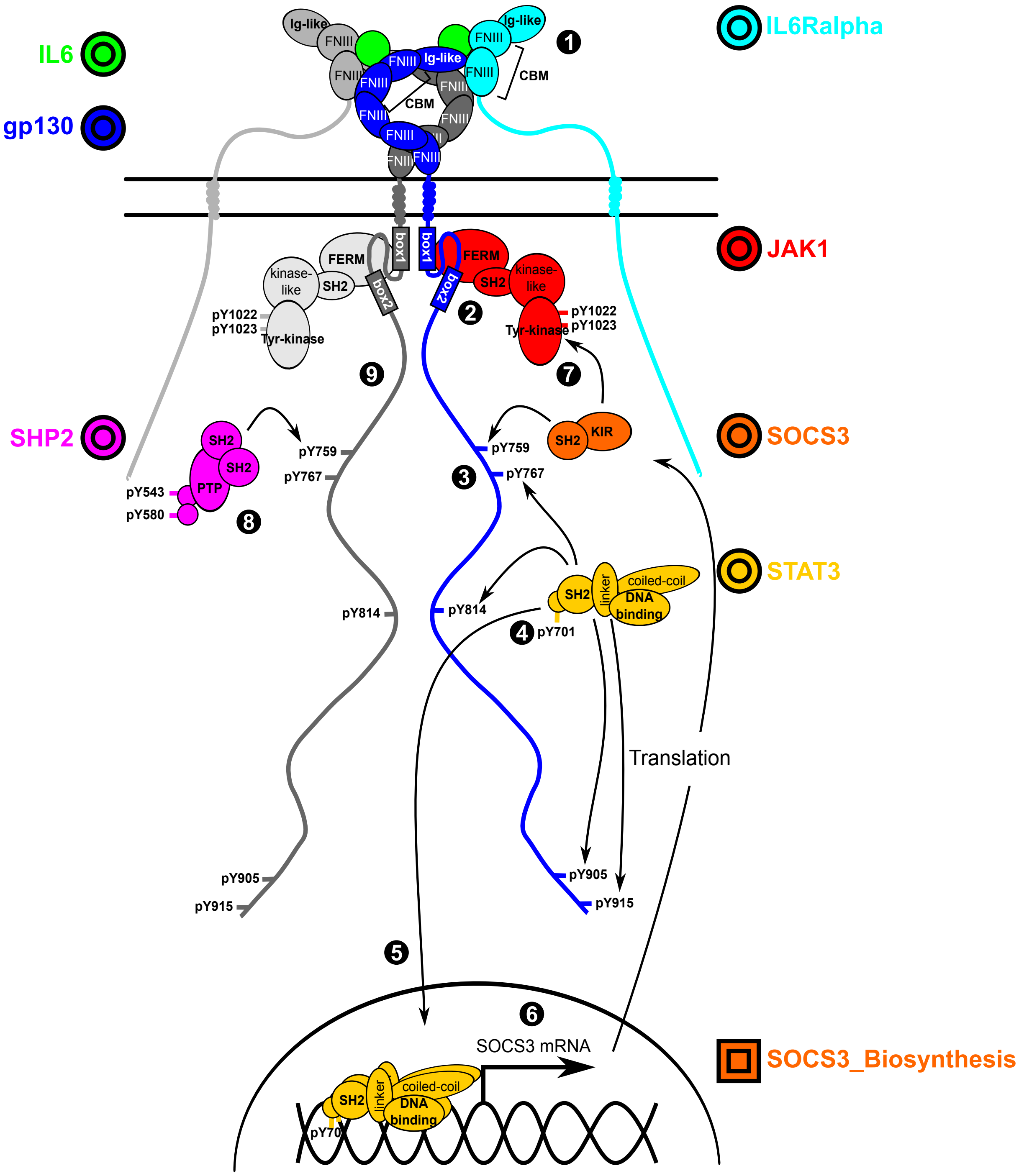

**Figure 2 Molecular model of IL-6 signalling through the JAK/STAT signal transduction pathway.** The cytokine interleukin-6 (IL-6) signals through a type I cytokine transmembrane receptor complex composed of the ligand binding IL-6 receptor-α chain (IL-6Rα) and the signal transducing glycoprotein gp130. Binding of IL-6 to its receptor IL-6Rα and to gp130 (1) causes the dimerization of the IL6Rα/gp130 receptor complex which leads to the activation of the JAK kinase by transphosphorylation (2). JAK remains constitutively bound to gp130 in both, its active and its inactive form. Active JAK phosphorylates several tyrosine residues of the cytosolic part of gp130 (3). The STAT transcription factor binds to phosphotyrosines of gp130 and is subsequently phosphorylated by active JAK (4). Phosphorylated STAT proteins dimerize and, as dimers, translocate into the nucleus (5) to activate the transcription of multiple genes including SOCS (6). SOCS acts as a negative feedback regulator of JAK in binding to phosphorylated Y759 of gp130 (7) causing the inactivation of the JAK kinase and in turn decreasing the rate of STAT3 phosphorylation. The SHP2 phosphatase counteracts JAK by dephosphorylating phosphotyrosines of gp130 (8) while JAK inactivates SHP2 by phosphorylating two of its tyrosine residues (9) forming a second negative feedback loop in this network which maintains a buffered basal level of JAK activity, STAT3 phosphorylation, and SOCS3 concentration even in the absence of IL-6.

The molecular mechanisms described in this scheme have been modeled in the form of one separate Petri net (protein module) for each of the involved proteins as indicated by a corresponding coarse place (◎) representing the underlying Petri net.

The synthesis and degradation of SOCS3 are modeled within a biosynthesis/degradation module as indicated by the corresponding coarse transition (⧈) representing the underlying Petri net. For the definition and the use of coarse places and coarse transitions for modelling and simulation see legend to Figure 1D. The structural model of the signalling IL-6 receptor complex was redrawn according to [8] where data for the naming and function of subunits are given.

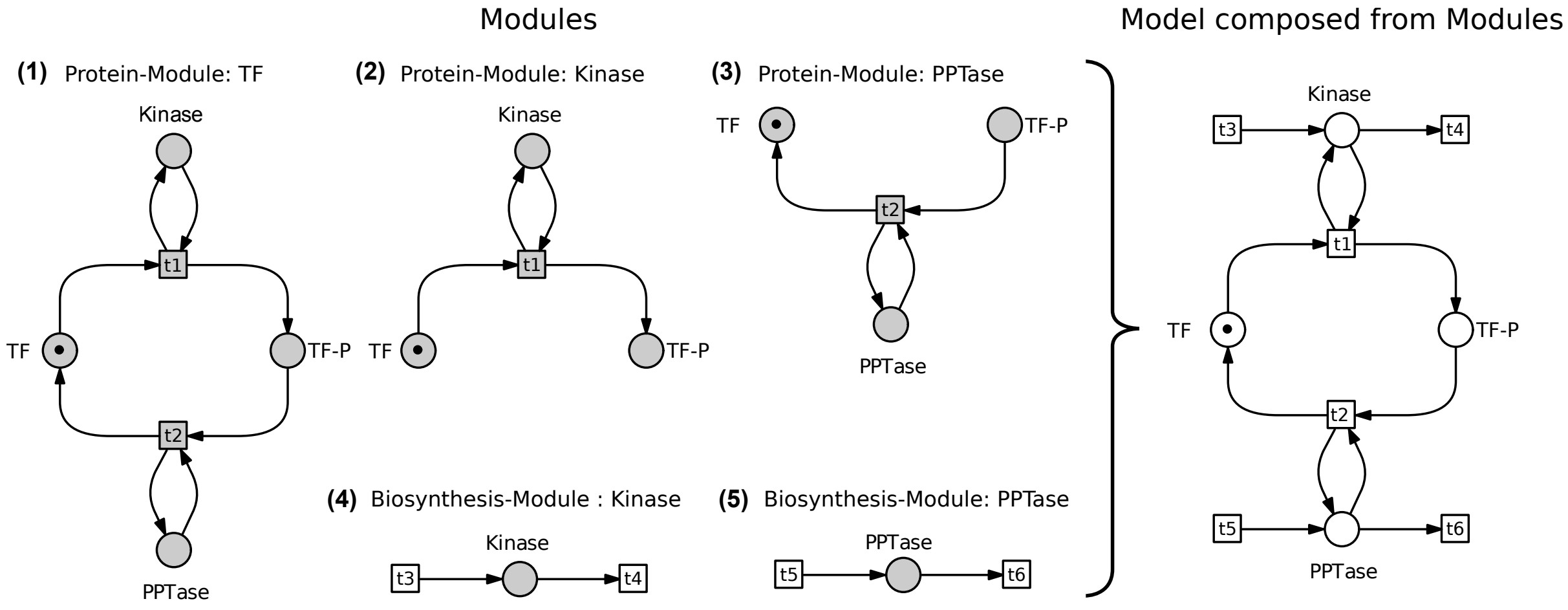

**Figure 3 Module types and design principles for composable Petri nets.** The scheme shows the composition of the Petri net of Figure 1E by combining three protein modules and two biosynthesis/degradation modules. Protein modules are bordered by places while biosynthesis/degradation modules are bordered by transitions. The modules represent the reactions of the transcription factor TF (1), the kinase (2), the phosphatase (3) as well as the metabolic turnover of kinase (4) and phosphatase (5), respectively. Places and transitions that are shared between the modules and accordingly define their interfaces are defined as logical nodes, i.e. as multiple graphical copies of the shared nodes. Considering the simple Petri net shown, representation in the form of individual modules appears as unnecessarily complicated as resulting modules may consist mainly of shared nodes. In modules of signal transduction networks, however, the benefits prevail.

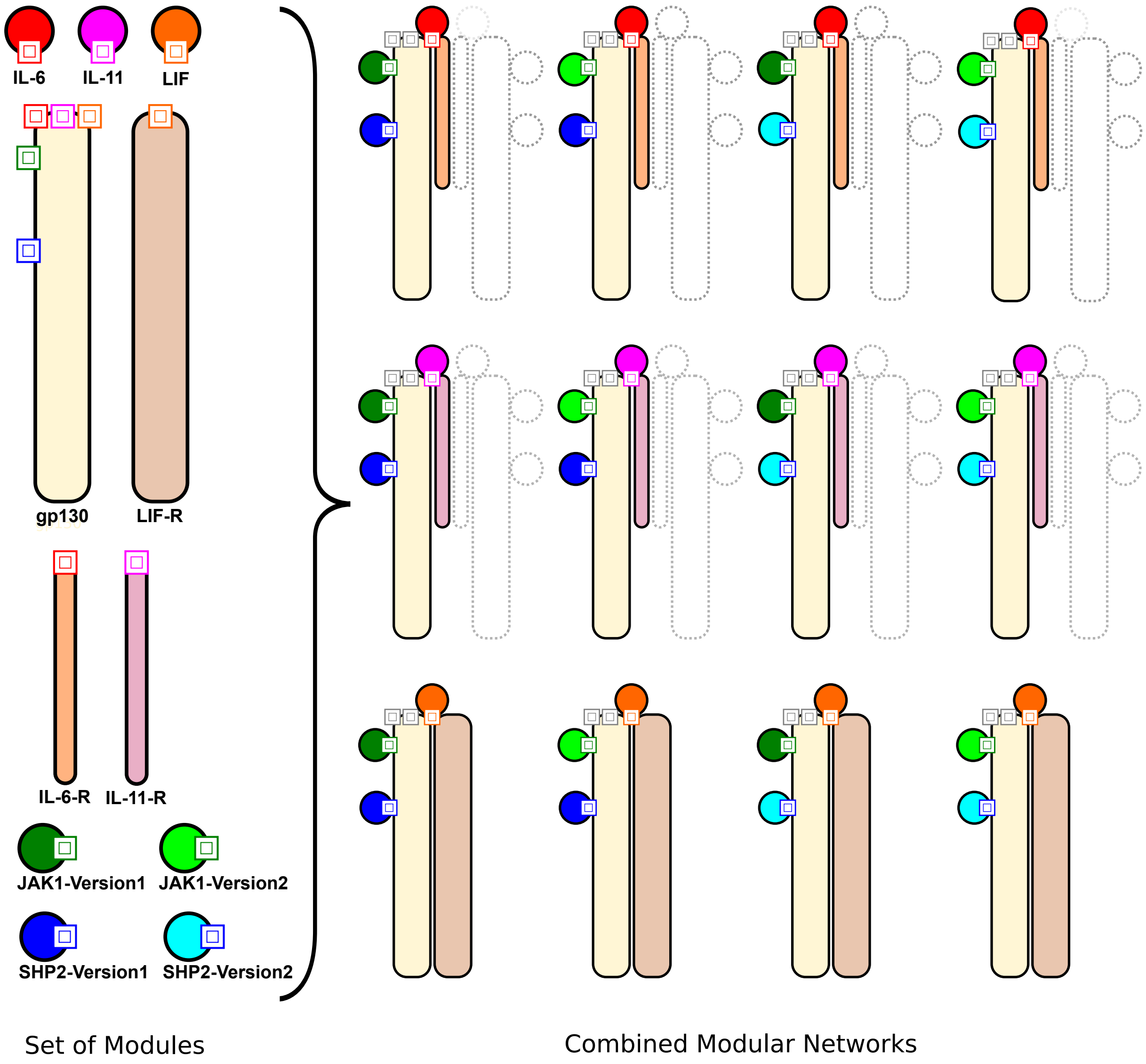

**Figure 4 Free combination of model modules for the automatic generation of composed models according to alternative reaction mechanisms or cell type-specific gene expression patterns.** The figure shows how protein modules of different cytokines, transmembrane receptors and modules representing alternative reaction mechanisms of JAK1 and SHP2 (left side) may be freely combined into alternative Petri net models of reaction networks of IL-6, IL-11, or LIF signal transduction or any combination thereof (right side). Automatic composition is mediated through networks of logical nodes (logical places and logical transitions)

organized into subnetworks each represented in the form of a coarse transition (⧈) that forms the specific functional interface between two given modules.

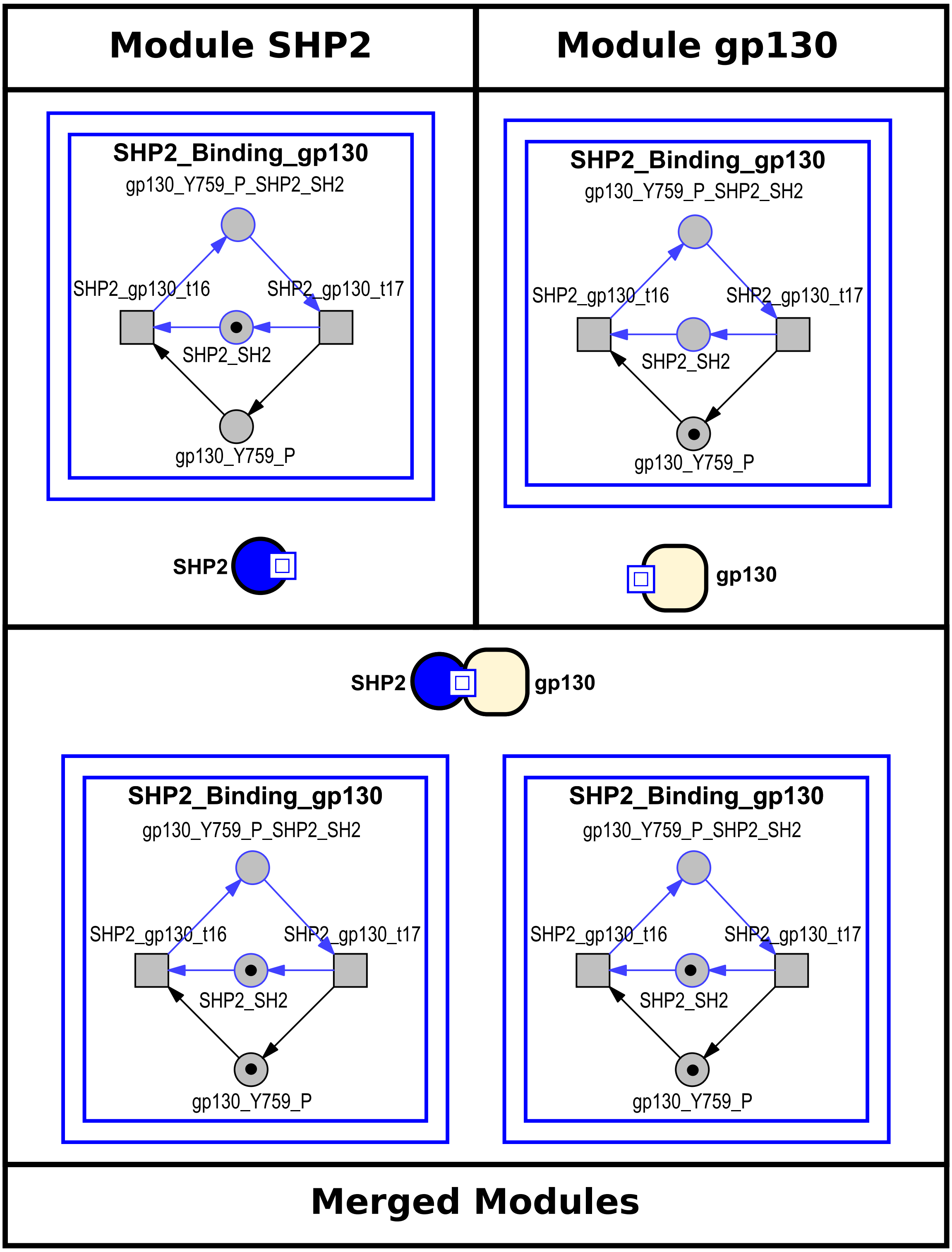

**Figure 5 The function of interface networks in connecting two modules depends on the marking.** In the example shown, identical copies of the interface network for connecting the SHP2 module and the gp 130 module are part of each of the two modules but their marking is different and the transitions that may functionally couple the two modules cannot fire unless the two modules are composed. Upon composition, the coupling transitions become automatically active.

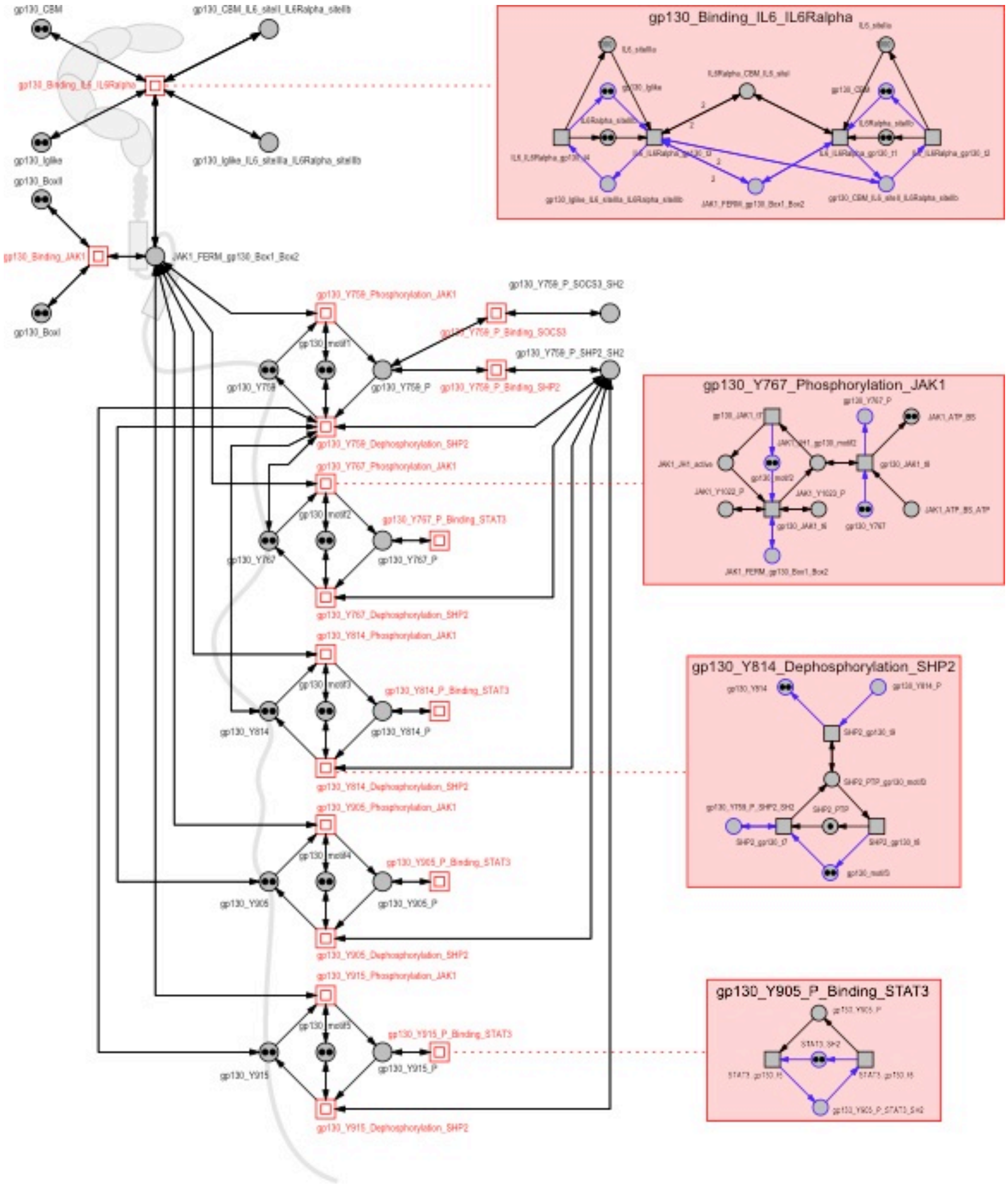

**Figure 6 The gp130 module.** The gp130 transmembrane receptor is part of the IL6-signalling complex. At the extracellular part the gp130 glycoprotein consists of the Ig-like domain and a cytokine-binding module. Both domains are responsible to form a complex with IL-6 and IL6-R. The intracellular part of gp130 has an interbox 1-2

region, where JAKs are constitutively bound. Downstream of this binding site the gp130 receptor has five important tyrosine residues (Y759, Y767, Y814, Y905, Y915), which are phosphorylated by activated JAK. The phosphotyrosine pY759 is the specific binding site of SHP2 and SOCS isoforms via their SH2 domain. Where STAT isoforms can interact via their SH2 domain with the other phosphotyrosines. In the corresponding Petri net model of the gp130 protein, the extracellular binding site of IL-6 and the intracellular binding site of JAK1 to the Box 1 and Box 2 sites are respresented as coarse transitions. Coarse transitions are also used to represent the phosphorylation and dephosphorylation reactions of the five tyrosins by JAK1 and SHP2, respectively. Specifically, three coarse transitions are assigned to each tyrosine residue downstream of Y759 describing the phosphorylation mechanism by JAK1, the dephosphorylation event by SHP2, and the binding interaction with STAT3. To the right of the coarse transitions representing the phosphorylation and dephosphorylation of Y759 (upper part of the figure), the Y759-P place is also connected with two coarse transitions describing the binding interaction with SHP2 and SOCS3. The boxed panels on the right side of the figure exemplify four subnets that are included in the corresponding coarse transitions in the gp130 module. Coarse transitions within the module are used for two reasons. (1) By coarse transitions, submodels encoding trivial and recurring mechanisms can be hidden. Hiding of those details is in favour of a clear layout of the important functional elements of the module that are arranged as much as possible according to the corresponding structural features of the protein. In the example given, the elements representing the reversible phosphorylation reactions of the tyrosines are clearly arranged in the sequence in which the tyrosines occur in the cytoplasmic domain of the protein (compare Figure 2). Accordingly, the sites of JAK binding and the binding sites of the ligand to the extracellular domain of the receptor have a obvious structural correlate in the Petri net. (2) Petri nets representing recurring mechanisms (e.g. phosphorylation/dephosphorylation reactions) hidden in a coarse transition can be conveniently cloned by copy/paste. This facilitates the editing of large networks and reduces the risk of modelling mistakes.

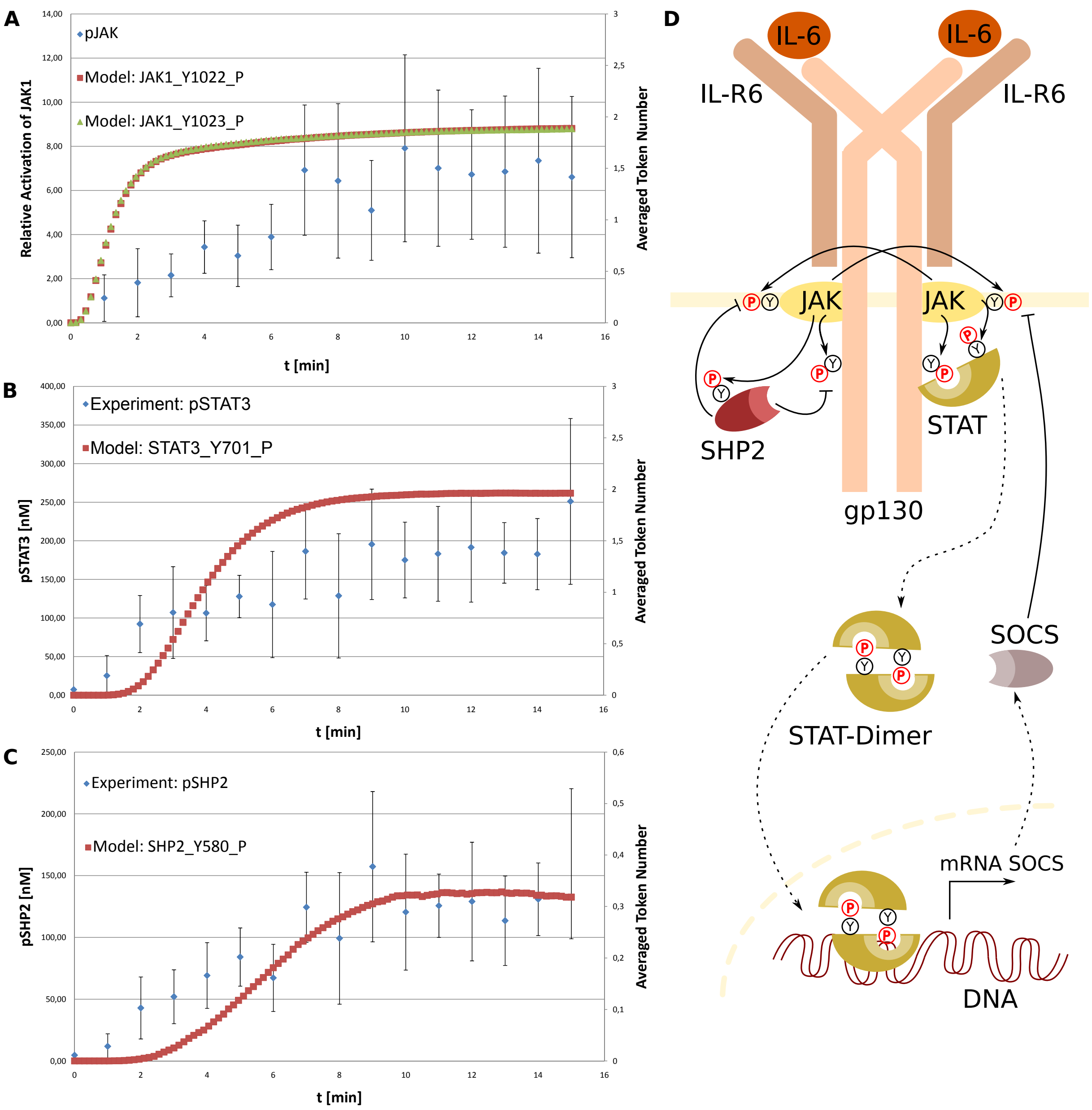

**Figure 7 Model of IL-6-induced JAK/STAT signalling assembled from the set of modules listed in Table 2 and tested against experimental data.** The model comprises the stepwise binding of IL-6 to its two receptor subunits IL-6R and gp130 and the subsequent activation of JAKs and its feed-back control. Activated JAKs phosphorylate each other as well as 5 different tyrosine residues of the cytoplasmic part of gp130 (Y759, Y767, Y814, Y905, Y915). The phosphorylated tyrosine residues serve as recruiting motif for monomeric STAT factors. Upon binding to the

receptor, STAT factors are phosphorylated by JAKs and dimerize via interaction of their SH2 domain with phosphorylated tyrosine residues. Dimerized STATs translocate to the nucleus were they induce the transcription of the feedback inhibitor SOCS3. The JAK/STAT pathway is regulated by the kinase inhibitor SOCS3 and the phosphatase SHP2. Both inhibitors bind to phosphorylated tyrosine 759 in gp130. Upon binding to the receptor SHP2 is phosphorylated by the JAKs. The Petri net model composed of modules represents the detailed kinetic mechanisms of the interaction between different protein domains (e.g. the assembly of the active receptor complex) and individual phosphorylation events at different tyrosine residues in the respective proteins (e.g. gp130). With the help of the biosynthesis/degradation modules the model also represents transcriptional activation, translation and protein decay of the feedback regulatory protein SOCS3, but at a more coarse grained level. To collect experimental data, human embryonic kidney cells (HEK293) stably expressing the human IL-6 receptor α (gp80) were stimulated with a 6 minutes pulse of IL-6 (20 ng/ml). Cells were harvested at the indicated time points and subsequently analyzed by immunoblotting with specific antibodies against (p)STAT3, (p)JAK1, (p)SHP2, and HSP70. The detection of HSP70 served as loading control. Data are presented as mean ± standard deviation from n= 5 to 10 independent experiments. Absolute concentrations of (p)STAT3 and (p)SHP2 were analyzed by quantitative immunoprecipitation using recombinant STAT3 and SHP2 proteins as calibrators. See (Dittrich et al., 2012, submitted) for original data.

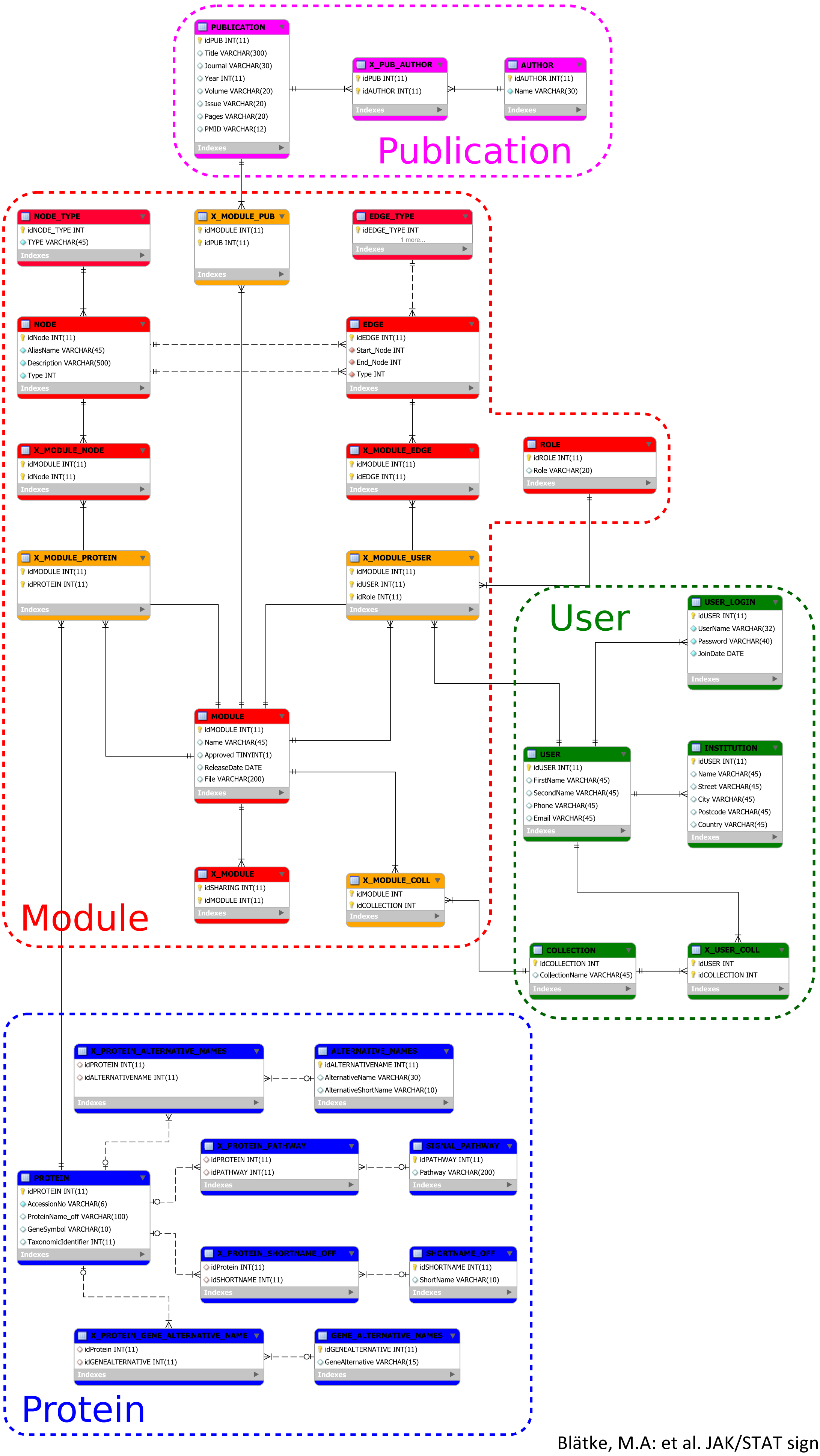

**Figure 8 Structure of the MySQL database.** The database management system used to establish the model module database is MySQL. The enhanced entity relation model (EER model) of the database consist of four functional parts containing the relevant datasets and supplementary information concerning the modules (red), proteins (blue), literature (pink), and users (green). The core part is the module scheme, which contains tables holding all information directly related to each of the modules and junction tables linking each module via its unique id with the corresponding protein, related literature, and user information (tables are indicated in the shade of the linked scheme). The information about each module is spread over three main tables, one holding general information about the module (id, name, state of approvement, release data, and file name), the other holding information about transitions and places (id, name and description). Junction tables link modules with their places and transitions. Another table indicates, which modules can be linked with each other. The tables of the protein schemes comprise the names assigned to each protein record (protein names, short names, gene names, synonyms), as well as the accession number, taxonomic identifier and a pathway description. All tables are again linked via junction tables. All information included in the protein schemes are extracted from the PubMed database using the unique accession number of each protein. The literature scheme stores information about the publications used to reference each module. The PMID is used to extract and reference publications in PubMed. The user scheme stores information, contact data, and affiliation of users as well as login data.

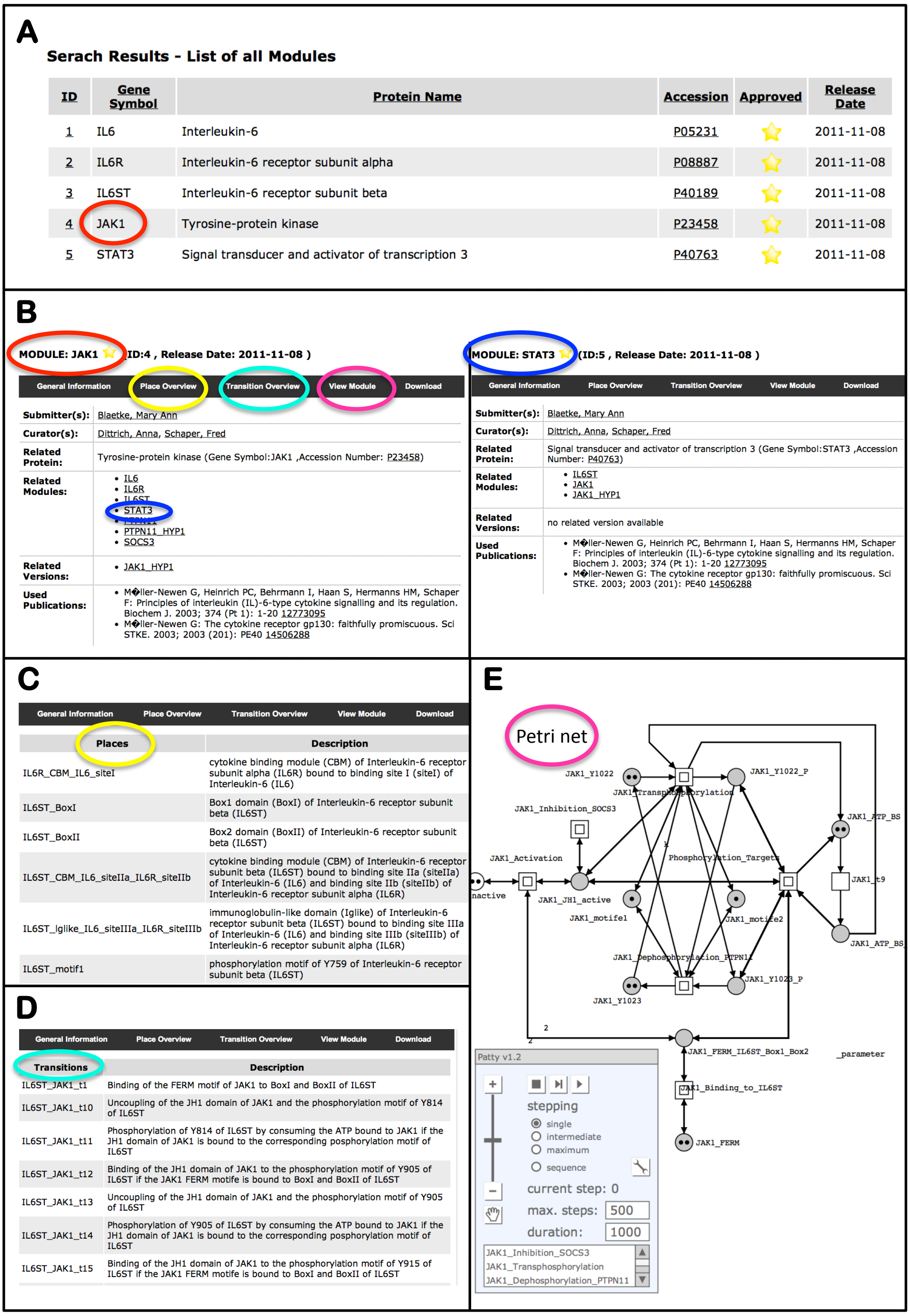

**Figure 9 Selcection of screenshots illustrating options of navigating through the searchable model module database.** (A) The modules contained in the module

base are listed according to the name of the protein, the gene symbol and the accession number linked to the respective dataset of the UniprotKB protein knowledgebase. By clicking on the gene symbol of a module, for example, the user obtains the module description page (B). Among other information, the page lists the complete set of interfaces to other modules of the database. By clicking on checkboxes, one can select with which other modules the current module should be coupled for generating a composed model. The list also contains all versions of connectable modules for a protein that might be in the database. By clicking on one of these names, the linked module description page opens (shown for STAT3) and displays with which proteins the module can be choosen to be connected. From the module description page, the user can access lists with the names of all places (C) or all transitions (D) or open the graphical layout of the Petri net (E) with a control window for the web-based animation of the token flow through the net. A preliminary implementation of the database is publically available at www.protbricks.de.

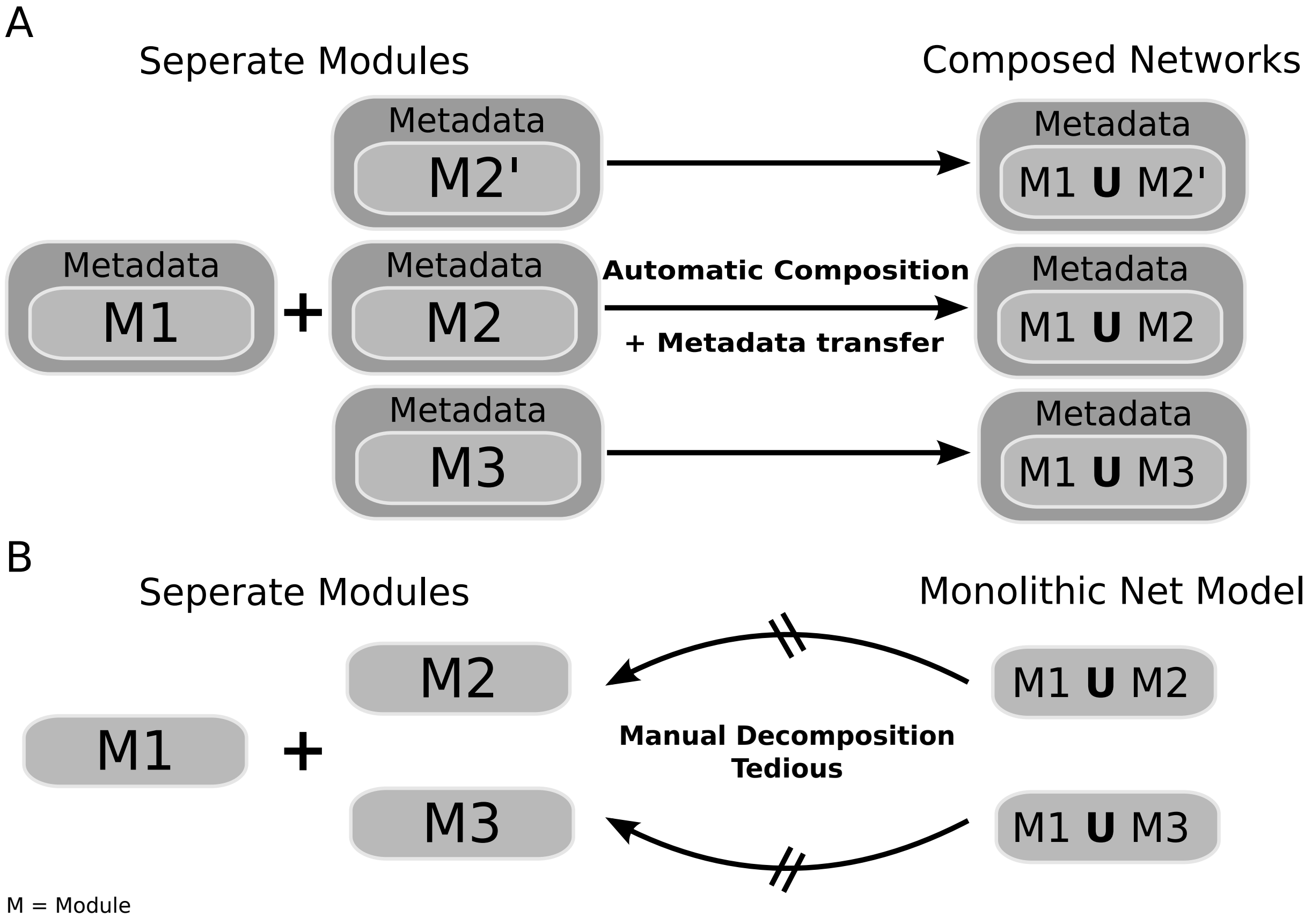

**Figure 10 Composition of models from modules can be done automatically and preserves metadata.** (A) Modules may be composed to multiple, well documented models, and updated models generated automatically from updated modules. (B) Manual decomposition of monolithic models in order to reuse parts of a model may be a considerable effort.

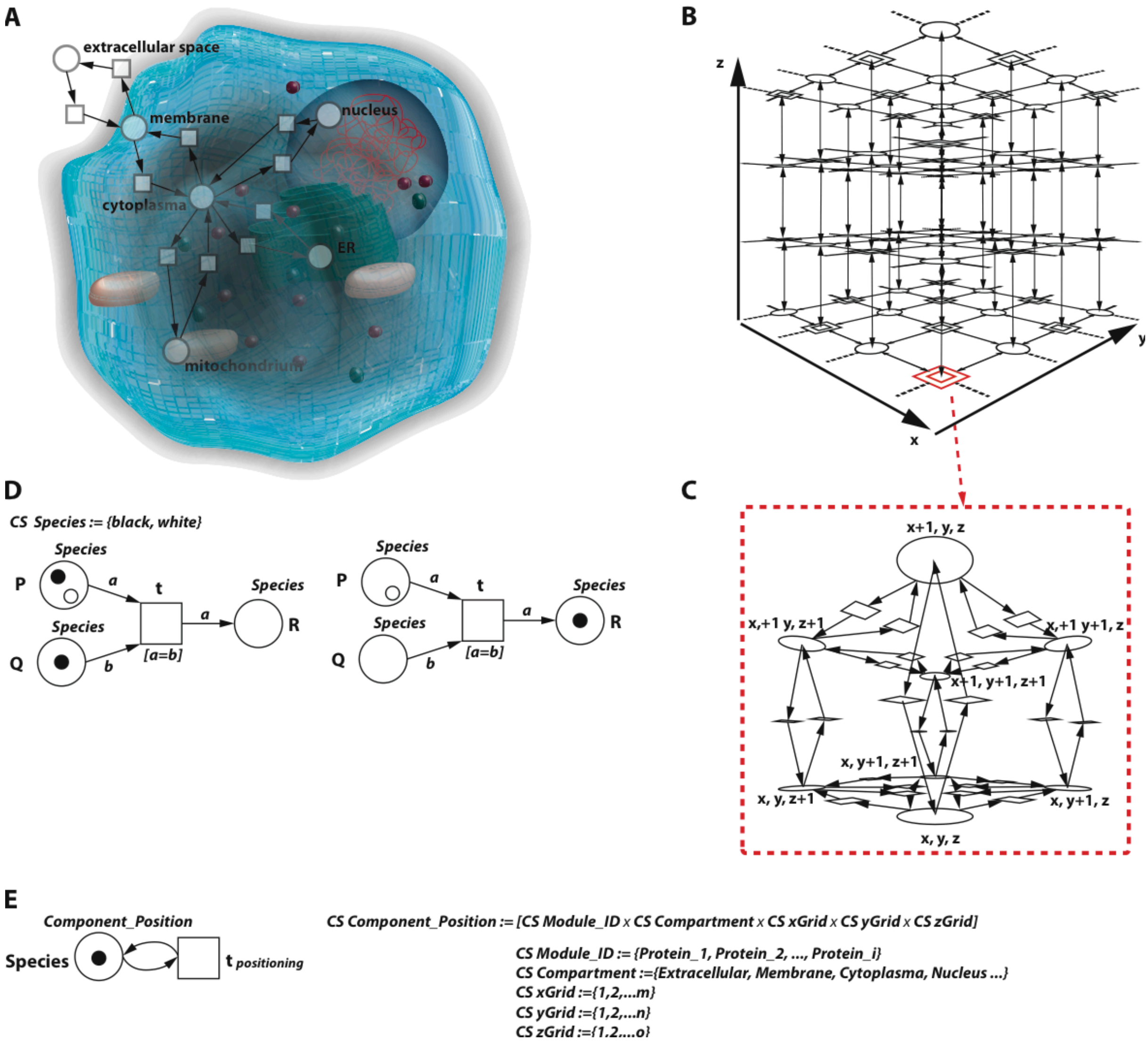

**Figure 11 A localisation component for the simulation of Petri nets with spatial resolution.** Interpretation of a model assembled from modules in the form of a coloured Petri net allows to perform simulations with spatial resolution and the representation of localisation in terms of cellular compartments with the help of a localisation component. A) The localisation of a molecule within the cell is defined by a token marking one place in a Petri net where places are assigned to functional compartments of a living cell (cytoplasm, cell membrane, mitochondrium, nucleus etc.). The net contains only one token for each module and is completely covered by one P-invariant. B) In order to add an arbitrarily fine three-dimensional spatial

resolution, one can design a Petri net component in the form of a three-dimensional lattice where each place corresponds to one position in space and a reaction volume defined by its coordinates (x,y,z). Again, the net is coverd by a single P-invariant and the spatial position of any molecule is unequivocally defined by a corresponding token in the net. The free diffusion of a molecule in the three spatial dimensions is simulated by the token changing between adjacent places through randomly firing transitions shown in (C). In coloured Petri nets, the three-dimensional lattice can be folded into one place and one transition (D). This transition is programmed in a way that each firing event of the transition updates the (x,y,z) coordinates encoded by the colour of the token. If the three dimensional lattice (B) was fitted into the 3-D topological model of a cell, the randomly moving token could get the cellular compartment as additional attribute and the transition between different cellular compartments can be eventually restricted according to this attribute. In the world of coloured Petri nets a token can update its colour according to its actual position while moving through the localisation component. In reality, two molecules can only react with each other if they reside in the same reaction volume. In the Petri net model of a biochemical reaction network, the localisation of a molecule in terms of volume element and cellular compartment can be encoded by the colour of the token which marks the place representing the corresponding molecule. In the coloured Petri net of panel (D), the firing rules of transition t are defined in a way that t takes only tokens of the same colour and preserves the colour for the token delivered into place R. The panel shows the Petri net before (left) and after (right) firing of transition t. E) The colour of the token may be defined as a tuple of five variables (Module_ID, compartment, x,y,z). In this case, the Petri net representing the localisation component can be folded into one place coupled to one transition.

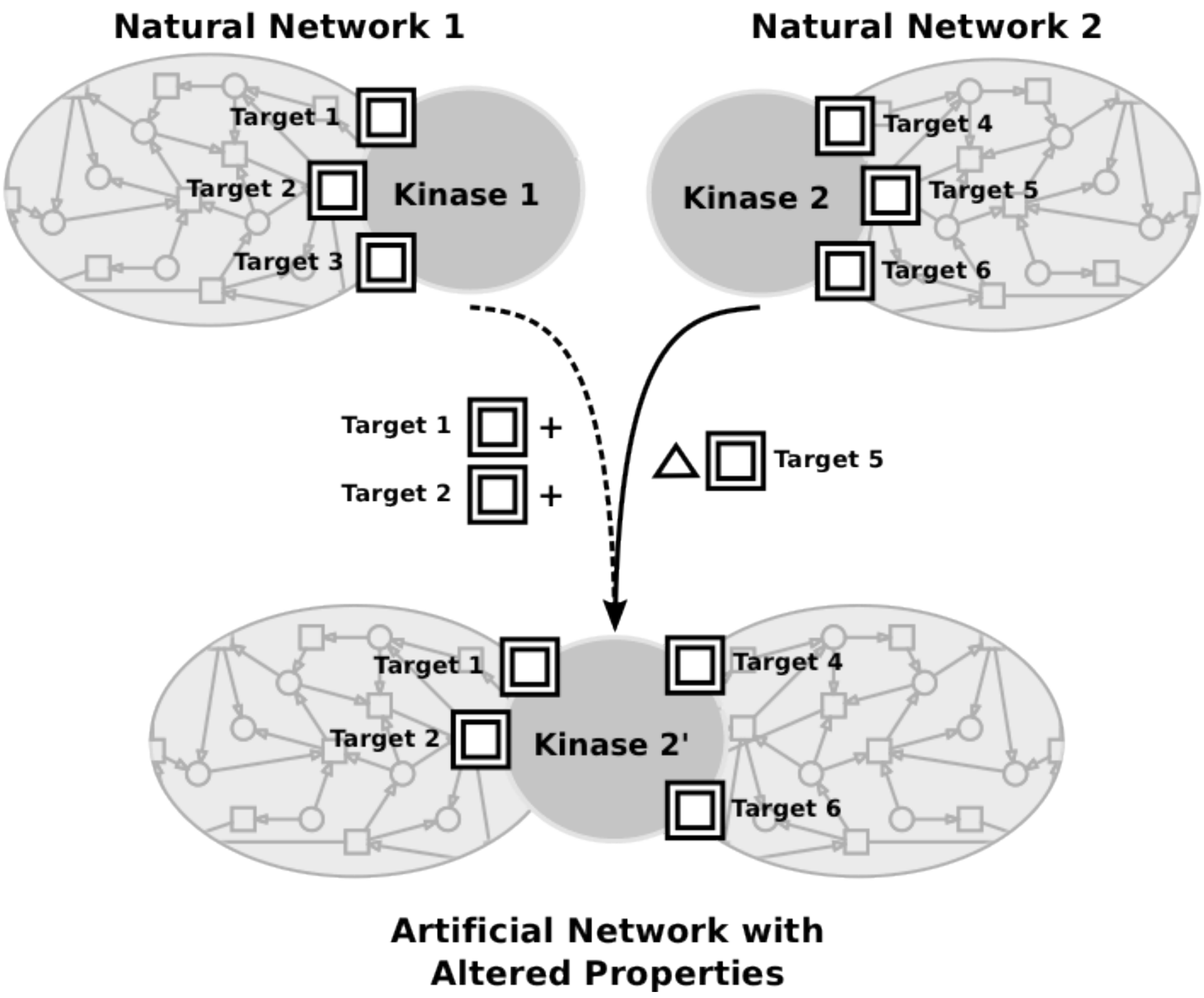

**Figure 12 Automatic biomodel engineering of synthetic or synthetically rewired networks by combination of modules representing altered binding sites.** Reengineering of modules by deleting existing and introducing new binding sites specific to other modules would allow to assemble *in silico* networks with alternative wiring that could be queried for pre-defined properties. In principle, the approach could be done fully automatically and systematically by making essential use of the metadata assigned to each module in order to avoid combinatorial explosion by sticking to biochemically realistic scenarios.